\documentclass{Interspeech2024}

\interspeechcameraready

\title{DeSTA: Enhancing Speech Language Models through Descriptive Speech-Text Alignment}

\name[affiliation={1}]{Ke-Han}{Lu}
\name[affiliation={2}]{Zhehuai}{Chen}
\name[affiliation={2}]{Szu-Wei}{Fu}
\name[affiliation={2}]{He}{Huang}
\name[affiliation={2}]{Boris}{Ginsburg}
\name[affiliation={2}]{Yu-Chiang Frank}{Wang}
\name[affiliation={1}]{Hung-yi}{Lee}

\address{
  $^1$Graduate Institute of Communication Engineering, National Taiwan University, Taiwan\\
  $^2$NVIDIA}
\email{\{d12942024, hungyilee\}@ntu.edu.tw, \{zhehuaic,szuweif,heh,bginsburg,frankwang\}@nvidia.com}


\usepackage{xcolor,colortbl}
\definecolor{Gray}{gray}{0.95}
\newcolumntype{C}{>{\columncolor{Gray}}c}

\usepackage{hyperref}       
\usepackage{url}            
\usepackage{booktabs}       
\usepackage{amsfonts}       
\usepackage{xcolor}         
\usepackage{graphicx}
\usepackage{amsmath,amssymb}
\usepackage{xurl}
\usepackage{multirow}
\usepackage{tabularx}
\usepackage{booktabs}
\usepackage{makecell}
\usepackage{float}
\hypersetup{colorlinks=true}
\usepackage{caption}
\usepackage{subcaption}
\usepackage{cite}

\begin{document}

\maketitle

\begin{abstract}

Recent speech language models (SLMs) typically incorporate pre-trained speech models to extend the capabilities from large language models (LLMs). In this paper, we propose a \textbf{De}scriptive \textbf{S}peech-\textbf{T}ext \textbf{A}lignment approach that leverages \textit{speech captioning} to bridge the gap between speech and text modalities, enabling SLMs to interpret and generate comprehensive natural language descriptions, thereby facilitating the capability to understand both linguistic and non-linguistic features in speech.
Enhanced with the proposed approach, our model demonstrates superior performance on the Dynamic-SUPERB benchmark, particularly in generalizing to unseen tasks. Moreover, we discover that the aligned model exhibits a zero-shot instruction-following capability without explicit speech instruction tuning. These findings highlight the potential to reshape instruction-following SLMs by incorporating rich, descriptive speech captions.
\footnote{\url{https://github.com/kehanlu/Nemo/tree/desta/examples/multimodal/DeSTA}}

\end{abstract}

\section{Introduction}

In recent years, the advent and continuous evolution of large language models (LLMs) have revolutionized the landscape of natural language processing (NLP), demonstrating remarkable performance across a diverse array of text generation and understanding tasks~\cite{achiam2023gpt, anil2023palm, touvron2023llama, touvron2023llama2, bai2023qwen}. The effectiveness of LLMs is significantly enhanced through the process of instruction tuning, which equips these models with the flexibility to adapt to novel tasks by following specific instructions~\cite{wei2022finetuned, zhang2023instruction}.  

\begin{figure}
    \centering
    \includegraphics[width=0.9\linewidth]{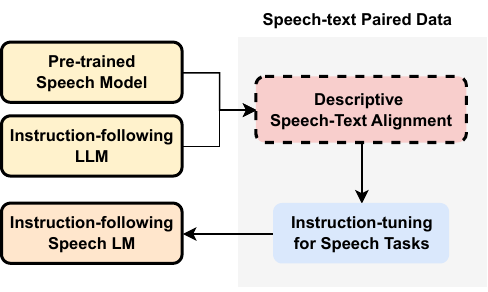}

    \caption{The Descriptive Speech-Text Alignment framework.}
    \label{fig:framework}
    \vspace{-1.3em}
\end{figure}

Motivated by the success of LLMs, recent studies have begun exploring the potential of LLMs in the realm of speech processing \cite{gong2023listen, gong_ltuas, tang2023salmonn, kong2024audio, shu2023llasm, wu2023decoder, deshmukh2024pengi, wang2023slm, pan2023cosmic, huang2023audiogpt}. These versatile instruction-following speech language models (SLMs) are designed to comprehend textual instructions and perform specific speech processing tasks. As depicted in Figure~\ref{fig:framework}, typically, these models incorporate a pre-trained speech model and instruction-following LLMs as fundamental architecture. 
Subsequently, these models undergo instruction-tuning across various speech tasks, aiming to harmonize the capabilities of the speech model with the text-based LLM.
Nevertheless, this process of aligning speech and text models presents multiple challenges. 
First, the necessity for curators to explicitly define the task scope is very difficult to standardize when assigning instructions across different tasks. It can also lead to overfitting problems~\cite{tang2023salmonn} to the training tasks and diminish the emergent capabilities of LLMs. 
Second, the training targets in current instruction-tuning datasets are usually formulated to perform one task at a time, which does not consider the multifaceted nature of speech. Finally, these datasets are usually formulated as classification tasks requiring only the model to predict one option~\cite{huang2023dynamic}. This process neglects the inherent long-form generation capability of LLM.

Figure~\ref{fig:framework} illustrates an overall framework in this paper. To address the highlighted issues, we present a novel descriptive speech-text alignment (DeSTA) stage before instruction tuning phase for SLMs. The speech-text-alignment stage aims at bridging the modality gap between speech and text through \textit{speech captioning}, thereby enables the broad understanding of speech. In this stage, we collect a speech caption dataset that features long-form natural language descriptions that encapsulate the multi-dimensional aspects of speech, by leveraging the metadata from existing datasets. This method mirrors the human ability to interpret and integrate multiple facets of speech into a comprehensive description. For example, \textit{"The woman shouted "I love cats" with joy, her voice loud, quick, and high-pitched..."} which captures not only the spoken words but also the speaking styles conveyed. To access the effectiveness of our speech-text alignment approach, we conduct instruction-tuning on our pre-trained model and evaluate our model on the Dynamic-SUPERB~\cite{huang2023dynamic} benchmark. 

\begin{figure*}[t]
  \centering
  \includegraphics[width=0.85\linewidth]{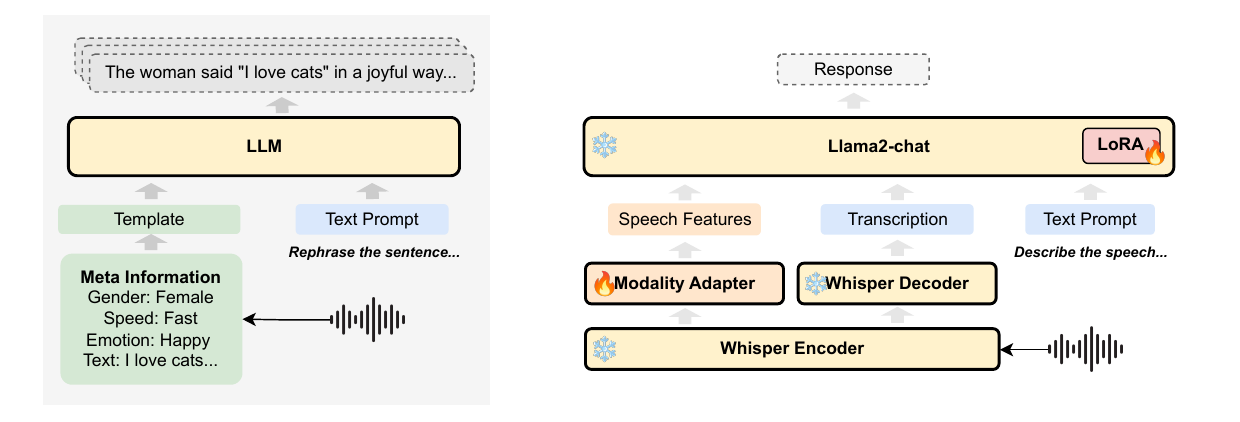}
  \caption{(Left) Data generation pipeline. (Right) Model architecture for speech-text alignment training.}
  \label{fig:model_and_data}
\end{figure*}

We summarize our contributions as follows.
\begin{itemize}
    \item We introduce a descriptive speech-text alignment training approach designed to train SLMs on both linguistic and non-linguistic information through a comprehensive speech caption. We will release our implementation and speech caption dataset in the future.
    \item With the proposed speech-text alignment training, our finetuned models outperform baselines systems on the Dynamic-SUPERB benchmark, particularly in tasks not covered during the training phase. This highlights the generalizability of our proposed framework.
    
    \item We discover that, although the pre-trained model is trained for speech captioning, it demonstrates zero-shot instruction-following capabilities derived from LLM by utilizing LoRA-scaling \cite{tang2023salmonn} at testing time.
\end{itemize}

\section{Related work}

Recently, there has been growing interest in integrating speech models with LLM to enable instruction-following for multitask speech processing. To this end, many recent innovations focus on creating speech instruction dataset that features a wide range of speech processing tasks for the instruction-tuning stage on Figure~\ref{fig:framework}. A common approach is to leverage the power of LLMs to automatically curate diverse instructions for speech tasks~\cite{shu2023llasm, gong2023listen, gong_ltuas, tang2023salmonn}. Although LLMs can not understand speech directly, it can understand and interpret the text description with strong knowledge of speech~\cite{gong2023listen}. For example, LTU~\cite{gong2023listen, gong_ltuas} generate large amounts of open-ended question-answering pairs from GPT. On the other hand, several studies decompose the learning process into multiple stage to enhance the learning efficiency and effectiveness~\cite{tang2023salmonn,chu2023qwen, liu2023music}. For instance, SALMONN~\cite{tang2023salmonn} presents a pre-training stage using speech recognition and audio captioning data and Qwen-audio~\cite{chu2023qwen} conduct multitask pre-training before instruction finetuning. 

\begin{table*}[th]
  \caption{Dynamic-SUPERB results. STA denotes the model is pre-trained with speech-text alignment. \textdagger We specify the seen and unseen tasks respect to the Dynamic-SUPERB training and evaluation set, while ASR+ChatGPT does not require training.}
  \label{tab:dynamic-superb}
  \footnotesize
  \setlength{\tabcolsep}{5.3pt}
  \centering
  \begin{tabular}{ l | c | c  c  c  c  c C | c  c  c  c c C | C}
    \toprule
    \multirow{2}{*}{\textbf{Model}} & & \multicolumn{6}{c|}{Seen} & \multicolumn{6}{c|}{Unseen} & \cellcolor{white}{All} \\
    
     & STA & \textbf{CON} & \textbf{SEM} & \textbf{PAR} & \textbf{DEG} & \textbf{SPK} & \textbf{Avg} &  \textbf{CON} & \textbf{SEM} & \textbf{PAR} & \textbf{DEG} & \textbf{SPK} & \textbf{Avg} & \textbf{Avg} \\
    \# Instances & & 9 & 2 & 4 & 6 & 3 & 24 & 2 & 4 & 3 & 13 & 2 & 24 & 48 \\
    \midrule
    ASR+ChatGPT\textdagger& &  67.11 & 43.75 & 38.75 & 47.50 & 35.67 & 52.78 & 50.50 & 71.63 & 5.17 & 40.31 & 47.75  & 42.60 & 47.10 \\
    ImageBind-LLM~\cite{huang2023dynamic} &  & 64.39&	48.25&	59.88&	79.17&	51.50 & 65.07 & 16.75&	22.50&	21.00&	48.65&	43.50 & 37.75 & 51.06 \\
    Whisper-LLM~\cite{huang2023dynamic}& &  76.94&	56.50&	68.00&	91.67&	92.17 & 79.28 & 8.00&	21.50&	6.83&	59.69& 59.25 & 42.38 & 60.85 \\
    \midrule
    CNN  &  & 90.33 & 63.25& 68.88& 49.33& 46.83& 70.70& 10.75& 50.25& 23.83& 48.00& 49.25& 42.35& 55.58 \\
    CNN  & \checkmark & 95.44 & 65.50& 63.13& 50.83& 57.83& 73.52& 71.25& 73.25& 11.00& 50.08& 45.00& 50.40 & 61.05 \\
    Qformer &  & 96.39& 65.00& 77.25& 44.33& 55.33 & 74.87 & 87.75& 80.50& 4.17 & 43.19& 46.25 & 48.50 & 60.47 \\
    Qformer & \checkmark & 95.00& 67.50& 74.38& 71.25& 59.00 & \textbf{80.15} & 74.50& 75.75& 20.33 & 57.54& 46.50 & \textbf{56.42} & \textbf{67.63} \\
    
    \bottomrule
  \end{tabular}
\end{table*}


\section{Descriptive Speech-Text Alignment}

In this work, we introduce a descriptive speech-text alignment training stage that aims at training the SLM to generate comprehensive speech captions. As demonstrate on Figure~\ref{fig:model_and_data}(Left), we utilize a large language model to generate speech captions with metadata from the audio. Then the SLM learns the multifaceted speech concept from the caption data accordingly. 

\subsection{Speech caption}

The primary objective of creating a speech caption dataset is to accurately capture the complex nature of speech and translate it into comprehensive, natural language descriptions. As demonstrated in Figure~\ref{fig:model_and_data}(Left), following the methodologies of \cite{gong_ltuas,gong2023listen}, we collect meta-information from existing datasets. This meta-information includes attributes such as speaking style (e.g., pitch, volume, and speaking speed), speaker information (e.g., gender), and the actual spoken content. We then empirically curate initial templates based on these attributes and formulate them into multiple natural language sentences. For example, a template might be structured as: \textit{"A [gender] speaker says [text] with [emotion] emotion."}
Next, we employ a LLM to generate diverse captions that reflect various writing styles and tones based on the provided sentences. Specifically, through prompt engineering, we instruct the LLM to accurately reflect the original spoken content while creatively incorporating the speech attributes to avoid hallucination. For each audio, we generate several captions based on different templates and prompts to ensure the dataset represents a wide range of expressiveness while avoiding repetition.

\subsection{Model architecture}

As depicted in Figure~\ref{fig:model_and_data}(Right), our architecture incorporates a pre-trained Whisper model~\cite{radford2022robust} alongside an instruction-following Llama2-chat model~\cite{touvron2023llama2}. Throughout the training phase, these pre-trained models remain unchanged.
A randomly initialized modality adapter is employed to map the speech feature from the Whisper encoder into the input representation space of Llama. Additionally, the transcribed text is fed into the language model as supplementary input. Low-rank adapters~\cite{hu2021lora} are attached to the LLM to enhance training efficiency.



\textbf{Modality adapter} The modality adapter is designed to extract meaningful representations from speech inputs. Specifically, the adapter processes the hidden outputs from the intermediate layers of the Whisper encoder to obtain high-level speech features. In this work, we explore two architectures: CNN and Qformer~\cite{pmlr-v202-li23q}. While the CNN preserves temporal information from the speech, the Qformer provides more flexibility with its attention mechanism. Next, these layer-wise representations are combined through a weighted summation using learnable weights to obtain the final representations. Finally, a projection layer is employed to map the continuous representations into the embedding space of Llama.

\textbf{Large language model} The speech features from the modality adapter and the transcribed text are concatenated with text prompts. In the speech-text alignment phase, we use a set of prompts for speech captioning, such as \textit{"Describe the speech"} and \textit{"What can be inferred from this audio?"}, to maintain the input structure of instruction tuning. As a result, the model is trained to generate speech captions with the next-token-prediction loss in this stage.

\section{Experiment}

\begin{table}[h]
\caption{Statistics of speech caption dataset.}
\footnotesize
\label{tab:pretrain_statistics}
\centering
\setlength{\tabcolsep}{5pt}
\begin{tabular}{l c c c}
\toprule
\textbf{Pretrain statics} & \textbf{LibriTTS} & \textbf{IEMOCAP} & \textbf{PromptTTS} \\
\midrule 
\# Audios & 20,807 & 4,262 & 23,544 \\
\# Captions & 62,309 & 12,781 & 70,439 \\
Duration(hours) & 88.3 & 16.2 & 132.6 \\
Avg. length(tokens) & 60.8 & 62.4 & 66.3 \\
\bottomrule
\end{tabular}
\end{table}

\subsection{Speech caption dataset}


In our study, we created a speech caption dataset by combining the LibriTTS~\cite{zen19_interspeech}, IEMOCAP~\cite{busso2008iemocap}, and PromptTTS~\cite{guo2023prompttts} datasets, all known for their expressive speech attributes. These datasets include meta information such as gender, pitch, volume, speaking speed, and text transcriptions. We utilized publicly available curated metadata~\cite{gong_ltuas} for LibriTTS and IEMOCAP, adding emotion labels for IEMOCAP. Additionally, we enriched our dataset with the PromptTTS dataset, which features emotional speech and human-annotated style description. We employed the Zephyr-7b-beta~\cite{tunstall2023zephyr}\footnote{{https://huggingface.co/HuggingFaceH4/zephyr-7b-beta}} to generate captions by randomly combining three prompts and five templates for each audio. 
Ultimately, as detailed in Table~\ref{tab:pretrain_statistics}, our dataset consists of 48,613 audio clips and provides 145,529 audio-caption pairs, totaling 237.1 hours, with an average length of 60 tokens each.

\subsection{Descriptive speech-text alignment}
We use the NeMo toolkits \cite{kuchaiev2019nemo} to implement the proposed method. We utilize publicly available checkpoints of Llama2-7b-chat\footnote{{https://huggingface.co/meta-llama/Llama-2-7b-chat}} and Whisper-large-v3\footnote{{https://huggingface.co/openai/whisper-large-v3}}, which have 1.5 billion parameters, as our fundamental architecture. The LoRA adapters (rank=32) are injected into the query, key, and value projection layers of the attention mechanisms within Llama. A scaling factor $\alpha$ is set to control the impact of these adapters at testing time~\cite{tang2023salmonn}. For processing speech inputs, the Whisper encoder generates 1,500 hidden representations for each audio. The Qformer architecture consists of a stack of two Transformer decoder blocks~\cite{vaswani2017attention}, coupled with 64 learnable query vectors. Similarly, for the CNN settings, two CNN layers are designed to downsample the encoder outputs into 60 representations for each audio. The speech is transcribed by the Whisper beforehand to enhance training efficiency. The total number of trainable parameters in our model is 56.3M for the Qformer architecture and 46.4M for the CNN architecture, respectively. We train the model on the speech caption dataset using the Adam optimizer with a cosine annealing scheduler for 5 epochs. We use 4 V100 GPUs and the global batch size is 12 with learning rate of 1e-4. 



\subsection{Instruction tuning}


The speech-text aligned model is further refined by utilizing the Dynamic-SUPERB~\cite{huang2023dynamic} training set. This dataset includes a wide range of instruction-guided speech processing tasks, which are categorized into five dimensions: content (CON), semantic (SEM), paralinguistic (PAR), degradation (DEG), and speaker (SPK). The training set comprises 22 instances, resulting in a total of 107K instruction pairs. Empirical evidence indicates that tasks within the content dimension are generally straightforward for LLMs to address. 
Therefore, we employ random selection to adjust our training data by incorporating fewer training samples from the content dimension, leading to a dataset with 76.5K samples. Additionally, we encountered an issue with task overfitting when employing a multi-task training approach. To mitigate this, we reduced the dataset size by 80\%, yielding a total of 15K data samples. Unless otherwise stated, we use this dataset configuration in our experiments. In the inference stage, we employ a reduced version of Dynamic-SUPERB evaluation set, containing 48 speech-related instances, each with 200 samples. We adhere to the evaluation settings to calculate the exact match accuracy for all instances. The complete task list is available at \url{https://github.com/dynamic-superb/dynamic-superb}.

\section{Results}

\subsection{Results on Dynamic-SUPERB benchmark}


Table~\ref{tab:dynamic-superb} presents the results from Dynamic-SUPERB, categorizing them into \textit{seen} and \textit{unseen} categories. The ASR+ChatGPT model is a cascaded system in which ChatGPT processes text transcribed by Whisper-large-v3. Both ImageBind-LLM and Whisper-LLM are baseline systems proposed alongside the Dynamic-SUPERB benchmark~\cite{huang2023dynamic}. ImageBind-LLM~\cite{han2023imagebind} is a multimodal language model that leverages an ImageBind~\cite{girdhar2023imagebind} encoder to integrate features from audio, vision, and text into a unified representation space. On the other hand, Whisper-LLM, based on the architecture of ImageBind-LLM, replaces the encoder with a Whisper encoder. Both models are further instruction-finetuned on the Dynamic-SUPERB training set, enabling them to follow instructions for executing speech processing tasks.



At first glance, the speech-text aligned Qformer model surpasses existing baseline systems, achieving an overall accuracy of 67.63\%. Notably, it excels in both seen categories and makes significant strides in the unseen category, where it achieve an average accuracy of 56.42\%. While CNN architectures are less effective than the Qformer, they exhibit superior performance in unseen categories compared to baseline models. It is important to highlight that there are no tasks explicitly related to the metadata employed during the speech-text alignment phase, with the exception of two emotion-related tasks in the seen category. 
Additionally, compared with the results to those without speech-text alignment, enhancements are evident in both seen and unseen categories. This indicates that the proposed method is not only good at adapting to new training tasks but also maintains the generalization capabilities of Llama.

Upon comparing performance across different dimensions, we observed that the Whisper-LLM significantly outperforms performance in the DEG and SPK dimensions. In contrast, our model demonstrates enhanced performance in the CON, SEM, and PAR dimensions. This disparity may be attributed to our method of incorporating transcribed text as input to the LLM, prompting the model to process both linguistic and non-linguistic features concurrently. However, We found that this may negatively impact the performance in tasks involving overlapping or noisy speech. As a consequence, our architecture faces challenges in such environments.

\begin{table}[t]
  \caption{Abalation studies on instruction-finetuning data size.}
  \label{tab:data_size}
  \footnotesize
  \begin{tabular}{c| cc | ccc}
  \toprule
  Model & STA & Data & Seen & Unseen & All \\
  \midrule
  ImageBind-LLM & & 107K & 65.07 & 37.75 & 51.06 \\
  Whisper-LLM & & 107K & 79.28 & 42.38 & 60.85\\
  \midrule
  \multirow{4}{*}{CNN} &  & 15K &70.70 & 42.35 & 55.58 \\
  & \checkmark & 15K & 73.52 & 50.40 & 61.05 \\
  & & 75.6K & 71.22 & 45.06 & 57.11 \\
  & \checkmark & 75.6K & 74.37 & 48.63 & 61.30 \\
  \midrule
  \multirow{4}{*}{Qformer} & & 15K & 74.87 & 48.50 & 60.47 \\
  & \checkmark & 15K & \textbf{80.15} & \textbf{56.42} & \textbf{67.63} \\
  & & 75.6K & 68.02 & 41.33 & 53.66 \\
  & \checkmark & 75.6K & 77.00 & 52.42 & 64.83 \\
  \bottomrule
  \end{tabular}
\end{table}

\subsection{Abalation studies on instruction-finetuning data size}
Table~\ref{tab:data_size} demonstrates the performance of Dynamic-SUPERB across different training sizes. Our model exhibits superior generalization capabilities on both seen and unseen categories when trained with 15K data points. Surprisingly, models trained with 75.6K data samples do not surpass the performance of those trained with a smaller dataset, despite the significant increase in data quantity. This phenomenon may be attributed to the structure of Dynamic-SUPERB, which is designed as a multiple-choice question task (i.e., a classification task with predefined options). Therefore, the model might readily identify patterns, which could lead to an overfitting issue, especially with specific choices such as \textit{'yes'} or \textit{'no'}. 

\subsection{Results on zero-shot instruction following}

We discovered that our models, pre-trained with speech-text alignment prior to instruction-based fine-tuning, exhibit the ability to follow instructions in a zero-shot manner. To demonstrate this zero-shot capability, we assessed the performance of the model, which had undergone only speech-text alignment training, by evaluating it on a set of 100 randomly chosen audio samples from the PromptTTS test dataset. We tasked our Qformer model with a variety of questions related to gender, pitch, volume, and speaking speed, using both multiple-choice and yes/no formats. Motivated by the findings in~\cite{tang2023salmonn}, we scaled LoRA factors to trade-off between textual knowledge and speech-text alignment learning. Table~\ref{tab:zero_shot} demonstrates the results, and Table~\ref{tab:zeroshot_examples} shows the generated examples based on different LoRA-scaling factors. Initially, the original model ($\alpha=1.0$), as expected, could not follow the instructions because it was trained for speech captioning. However, when we lowered the LoRA-scaling factor, the model began to follow instructions and responded in a more precise way. Surprisingly, even when we completely removed the LoRA adapter ($\alpha=0.0$), the model still performed well. This shows that the proposed speech-text alignment training can effectively bring speech models to the LLM modeling space through the learned modality adapter, and reducing the LoRA-scaling factor makes the LLM behave more like its original form (i.e., instruction-following LLM). These findings imply that with the proposed speech-text alignment approach, we can utilize the instruction-following ability derived from LLMs without explicitly curating the instruction set for specific tasks by applying pruning methods during the testing time.

\begin{table}[t]
  \footnotesize
  \caption{Zero-shot instruction following results based on different LoRA-scaling factor. Success rate denotes the model follow the instruction and respond correctly, while accuracy indicates the performance when model follows the instruction.}
  
  \label{tab:zero_shot}
  \centering
  \begin{tabular}{c ccccc}
    \toprule
    $\alpha$ & Success Rate & Accuracy & Following Rate \\
    
    \midrule
    $1.00$ & N/A & N/A & N/A \\
    $0.75$ & 38.25 & 71.00  & 53.88  \\
    $0.50$ & 69.75 & 70.72  & 98.63  \\
    $0.25$ & 68.83 & 68.63 & 100.00  \\
    $0.00$ & 62.63 & 62.63 & 100.00  \\
    \bottomrule
  \end{tabular}
\end{table}

\begin{table}[t]
  \caption{Generation examples from a sentence spoke by a \textbf{female} speaker in a \textbf{slow} pace.}
  
  \label{tab:zeroshot_examples}
  \centering
  \footnotesize
  \begin{tabular}{l l}
    \toprule
    $\alpha$ & Response \\
    \midrule
    \multicolumn{2}{l}{\textit{\makecell[l]{Transcription: "Of course not," said he, jumping up from the throne,\\ where he had seated himself.}}} \\
    \midrule
    
    \multicolumn{2}{l}{\textcolor{blue}{Q: What is the gender of the speaker? Answer 'male' or 'female'.}}\\
    1.00 & The speaker's voice is soft and gentle,... \textit{(Description)} \\
    0.75 & The speaker's gender is identified as female. \\
    0.50 & The speaker is a female. \\
    0.25 & Male \\
    0.00 & Male \\
    \midrule
    \multicolumn{2}{l}{\textcolor{blue}{Q: Does the speaker talk in a fast pace? Answer 'yes' or 'no'.}}\\
    1.00 & The speaker's volume is soft, which...\textit{(Description)}
    \\
    0.75 & \makecell[l]{ The speaker's voice is soft and low, with a slow pace that \\allows for clear enunciation of each word.} \\
    0.50 & No. \\
    0.25 & No. \\
    0.00 & No \\
    \bottomrule
  \end{tabular}
\end{table}

\section{Conclusion}



In conclusion, our study introduces a descriptive speech-text alignment approach by leveraging speech captioning. This method bridges the modality gap between speech and text, enabling SLMs to comprehend a wide range of speech features. The experimental results on the Dynamic-SUPERB benchmark and the zero-shot evaluation demonstrate the effectiveness of our approach, particularly in its ability to generalize to new tasks. This indicates the potential to harness the general capabilities from instruction-following LLMs with our approach.



\section{Acknowledgement}
We thank the National Center for High-performance Computing (NCHC) of National Applied Research Laboratories (NARLabs) in Taiwan for providing computational and storage resources.

\bibliographystyle{IEEEtran}
\bibliography{mybib}

\end{document}